# Far-field Imaging beyond Diffraction Limit Using Single Radar

Lianlin Li[1], Fang Li[2], Tiejun Cui[3], Yunhua Tan[1,], Kan Yao[1,]

[1]School of EECS, Peking University, Beijing, 100871, China
[2]Institute of Electronics, Chinese Academy of Sciences, Beijing, 100080, China
[3]State Key Laboratory of Millimeter Waves, Southeast University, Nanjing, 210096, China
[*]lianlin.li@pku.edu.cn

**Far-field imaging beyond the diffraction limit is a long sought-after goal in various imaging applications, which requires usually an array of antennas or mechanical scanning. Here, we propose to solve this challenging problem using a single radar system consisting of a spatial-temporal resonant aperture antenna (referred to as the slavery antenna) and a broadband horn antenna (referred to as the master antenna). We theoretically demonstrate that the resonant-aperture antenna is capable of converting part evanescent waves into propagating waves and delivering them to far fields. We show that three basic requirements should be satisfied on the proposed subwavelength-imaging strategy: the strong spatial-temporal dispersive aperture, near-field coupling, and temporal (or broadband) illumination. The imaging concept with single radar provides the unique ability to achieve super resolution for *real-time data* when illuminated by broadband electromagnetic waves, without the harsh requirements such as near- field scanning, mechanical scanning, or antenna arrays. We expect the imaging methodology to make breakthroughs in super-resolution imaging in microwave, terahertz, optical, and ultrasound regimes.**

Far-field imaging beyond the diffraction limit is a long sought-after goal across various imaging applications in the microwave, terahertz, optical and ultrasound regimes. One of the key issues for this challenging topic is to exploit, whether directly or indirectly, the evanescent waves emerged from the objects under consideration [1-5]. For instance, the super-oscillation imaging is a technique which relies on the strong coherence between the evanescent waves and propagating waves for the objects fallen into a very narrow field of view (FOV) [e.g., 3 and references therein]. Moreover, a cluster of scatters are introduced in the vicinity of the probed objects for converting the evanescent waves into propagating waves, leading to an image of probed objects with subwavelength resolution [4, 5].

Conventionally, there are two types of popular active imaging systems for data acquisition: the real-aperture (RA) system and synthetic-aperture (SA) system. The SA system relies on the mechanical movement of a single radar to form virtually a large scanning aperture via post-processing, which is typically inefficient in data acquisition [6]. On the contrary, the RA system is composed of a large number of antenna elements, which has more flexibility in measurement modes, but sacrifices the size, weight, power, and price advantages of the single radar system [7]. Now, a natural question arises: *is it faithful to get a subwavelength image from a single radar*? The answer is encouraging [8, 9, 10]. In Ref. [8], Pierrat et *al.* investigated theoretically the feasibility of subwavelength focusing of a single broadband dipole source in an open disordered medium with a single antenna. Inspired by the theory of compressive sensing (CS), Hunt et *al.* and Lipworth et *al.* developed the meta-imager consisting of a single antenna with multiple resonant frequencies, along with a sparsity-promoted nonlinear solver [9, 10], by which a *high-resolution* imaging of the scene with sparse objects is achievable.

In this work, we propose a subwavelength imaging theory using a single radar from the far-field measurements, which consists of a slavery antenna with the spatial-temporal resonant aperture and a mater broadband horn antenna, as sketched in Fig. 1. The slavery antenna consists of a homogeneous square aperture with finite extension $L_x \times L_y$ and thickness $d$, which is filled by dispersive materials with the relative permittivity described by the Drude model $\varepsilon_r = 1 - \frac{\omega_p^2}{\omega^2 - i\omega\Gamma}$, where the bulk plasmon frequency $\omega_p = 1.5 \times 10^{10} rad/s$, and the collision frequency $\Gamma = 7.73 \times 10^7 rad/s$ are considered in this work. In the following discussions, we demonstrate that such a resonant aperture manages the conversion of evanescent waves into propagating waves, and hence the single radar system is capable of producing super-resolution imaging without using any near-field scanning, mechanical movement, and antenna array.

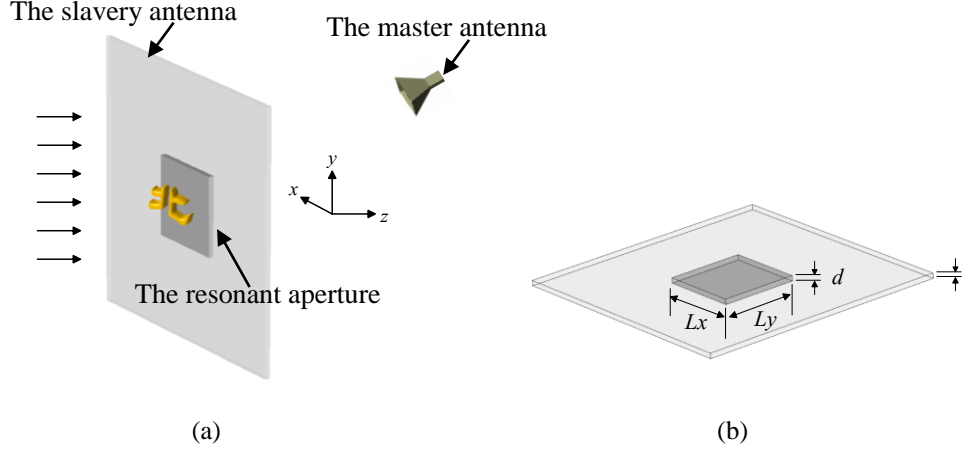

FIG.1 (a) The schematic setup of the single-radar imaging system for the far-field imaging beyond the subwavelength resolution, which consists of a mater antenna of broadband horn and a slavery antenna with resonant aperture. Here, the object under consideration is a Chinese word "北". (a) The schematic configuration of the slavery antenna, which is a square hole in a perfect conductor filled with a dispersive material characterized by the Drude model $\varepsilon_r = 1 - \frac{\omega_p^2}{\omega^2 - i\omega\Gamma}$, where $\omega_p = 1.5 \times 10^{10} rad/s$ and $\Gamma = 7.73 \times 10^7 rad/s$.

First, we present a theoretical analysis to provide the physical insights into the methodology of far-field imaging beyond the diffraction limit. Assuming that the resonant aperture is illuminated by a transverse-magnetic (TM) polarized plane wave $\boldsymbol{E}_{inc} = \boldsymbol{E}_0 e^{i\boldsymbol{k}_{in}\cdot\boldsymbol{r}}$, where $\boldsymbol{k}_{in} = (k_{in,x}, k_{in,y}, k_{in,z})$, $k_{in,z} = \sqrt{k_0^2 - k_{in,x}^2 - k_{in,y}^2}$, and $k_0$ is the free-space wavenumber. Note that the case of $k_{in,\rho} = \sqrt{k_{in,x}^2 + k_{in,y}^2} > k_0$ corresponds to the illumination of evanescent waves, and $k_{in,\rho} < k_0$ for propagating wave. In light of the vector Huygens' diffraction principle [9, 10], the electrical field perpendicular to the **z** direction at far field, $\boldsymbol{r}_d = (\boldsymbol{\rho}, z)$, amounts to be (see Appendix for details):

$$\boldsymbol{E}_t(\boldsymbol{r}_d; \boldsymbol{k}_{in}) \propto T^{TM}(\boldsymbol{k}_{in}) \times \text{sinc}\left(\frac{L_x(k_{r,x}-k_{in,x})}{2\pi}\right) \text{sinc}\left(\frac{L_y(k_{r,y}-k_{in,y})}{2\pi}\right) \quad (1)$$

where $k_{r,x} = k_0 \sin\theta_r \cos\varphi_r$, and $k_{r,y} = k_0 \sin\theta_r \sin\varphi_r$. Herein, $\theta_r$ and $\varphi_r$ are the polar and azimuth angles of $r_d$ in the spherical coordinate system, respectively. It is noted that $|T^{TM}(k_{in})|$ shows resonant peaks when the following condition is satisfied: $R_{op}^{TM} e^{ik_{pz}d} = 1$ ( $k_{pz} = \sqrt{\varepsilon_r k_0^2 - k_{in,x}^2 - k_{in,y}^2}$, and $k_{in,\rho} > k_0$ ).Such peaks are results of the Fabry-Perot resonance happened inside the resonant aperture, and can also be interpreted with the complicated interactions of surface plasmon polariton (SPP) modes in a truncated thin film. In other words, such phenomenon from the finite resonant aperture is related to the SPP scattering due to the geometric discontinuity [13].

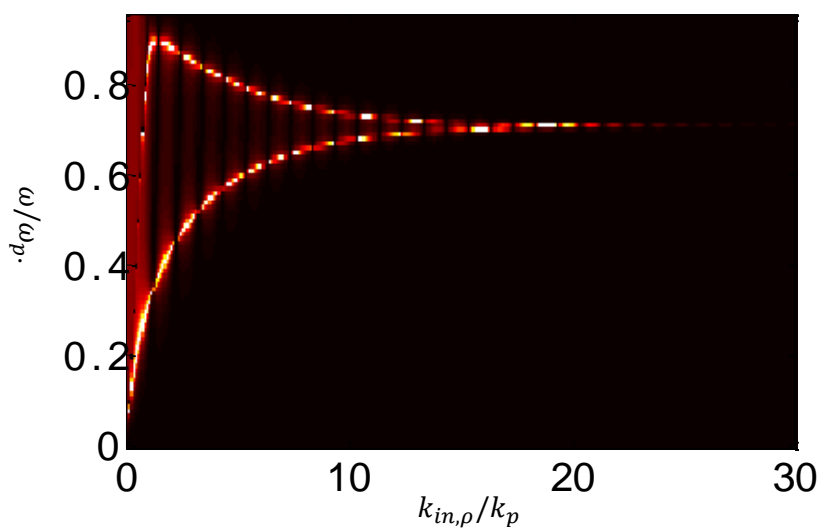

FIG. 2. The amplitudes of *x*-polarized electric field acquired at $r_d = (2.5\lambda_0, 5\lambda_0)$, where the incidence is a TM-polarized plane wave, the *x*-axis is $k_{in,\rho}/k_p$, and the *y*-axis denotes $\omega/\omega_p$. This result calculated by using Eq.(1).

A set of simulations are performed to verify the above claims. Figure 2 shows the dependence of the *x*-polarized electric field acquired at $r_d$ on $k_{in,\rho}/k_p$ ($k_p = 2\pi/\omega_p$, and $k_p d = 0.25$ is considered below) and $\omega/\omega_p$, which is calculated using Eq. (1). From these figures, we can see that the illumination of evanescent waves can be efficiently captured by the sensor at far fields, implying that the evanescent waves have been converted into propagating waves after experiencing the resonant aperture, as illustrated by the light region. Furthermore, the light region is associated with the dispersive property of the resonant aperture filled by the Drude medium, i.e.,

$$R_{op}^{TM} e^{ik_{pz}d} = \pm 1, \qquad (2)$$

which can be numerically solved to get the relation between $k_{in,\rho}/k_p$ and $\omega/\omega_p$. Since $R_{op}^{TM}$ is strongly subject to both the operational angular frequency $\omega$ and the incident wavenumber vector $\boldsymbol{k}_{in}$, we can deduce from Eq. (2) that this resonant aperture is of strongly spatial-temporal dispersive, and responsible for the efficient conversion between the evanescent waves and the propagating waves. In this way, the resonant aperture plays a role of resonant metalens [14]. Different from Ref. [14] where a series of far-field patterns are required to achieve a subwavelength imaging, the purpose of this work focuses on the feasibility of subwavelength imaging using a single sensor.

Now, we can confirm that part evanescent waves incident to the finite resonant aperture are converted into propagating waves. To show this more clearly, we make the following arguments. For an infinite homogeneous film, its eigenmodes are usually taken as $e^{ik_x x}$ whether inside or outside the film, which implies that there is no conversion between the evanescent and the propagating waves. However, the eigenfunction is not $e^{ik_x x}$ anymore when the film is truncated into a finite size in the lateral dimension, which means that the evanescent waves and propagating waves are now coupled inside the resonant aperture, as illustrated by the coupled-mode theory (For instance, Ref. [15]). More importantly, the propagating waves converted from the evanescent illumination can be significantly enhanced by the Fabry-Perot resonance happened in the resonant aperture filled with dispersive materials. In order to see this claim, a set of numerical simulations are conducted in the context of coupled-mode theory. We adopt the notations used in Ref. [15] to avoid introducing too many notations. The field emerged from the resonant aperture can be represented by

$$H(\mathbf{r}) = \sum_{m=0}^{Int(2L_x/\lambda)} E_m^O G_m(\mathbf{r}) + \sum_{m=Int(2L_x/\lambda)+1}^{\infty} E_m^O G_m(\mathbf{r}) \qquad (3)$$

Herein $\lambda$ is operational wavelength, and $Int(2L_x/\lambda)$ denotes the maximum integer less than $2L_x/\lambda$. In Eq. (3), the first term corresponds to the propagating waves emerged from the resonant aperture, while the second term for evanescent waves.

Figure 3 gives the map of $PS = \sqrt{\sum_{m=0}^{Int(2L_x/\lambda)} |E_m^O|^2}$, which describes roughly the energy of propagating waves emerged from the resonant aperture, as the function of on $k_{in,\rho}/k_p$ and $\omega/\omega_p$, where simulation parameters are same as those considered in Fig. 2. From this figure, we clearly see that the propagating waves can be generated and efficiently enhanced by the Fabry-Perot resonance happened in the resonant aperture.

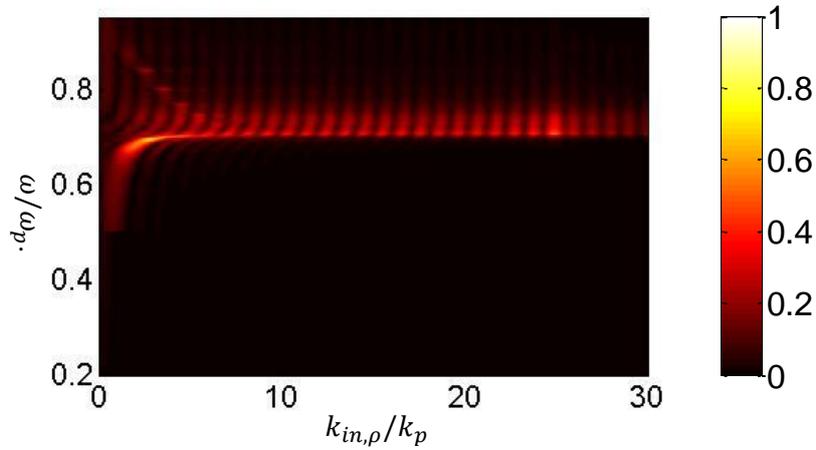

FIG. 3. The map of $PS = \sqrt{\sum_{m=0}^{Int(2L_x/\lambda)} |E_m^O|^2}$ normalized by its maximum as the function of $k_{in,\rho}/k_p$ and $\omega/\omega_p$, where simulation parameters are same as those considered in Fig. 2.

Second, the essential point of improving the resolution of an imaging system is to enhance its information capacity, which is usually represented by the temporal-bandwidth product [16, 17]. Here, we demonstrate that the information capacity of the proposed single radar imaging system can be efficiently driven up by the resonant aperture, in combination with the broadband illumination. From Eq. (1), we can derive the sensitivity of the frequency-dependent measurements with respect to frequency, namely,

$$\frac{d}{d\omega} \boldsymbol{E_t}(\boldsymbol{r}_d; \boldsymbol{k}_{in}) \propto \frac{d\,T^{TM}(\boldsymbol{k}_{in})}{d\omega} \times sinc\left(\frac{L_x(k_{r,x}-k_{in,x})}{2\pi}\right) sinc\left(\frac{L_y(k_{r,y}-k_{in,y})}{2\pi}\right). \qquad (4)$$

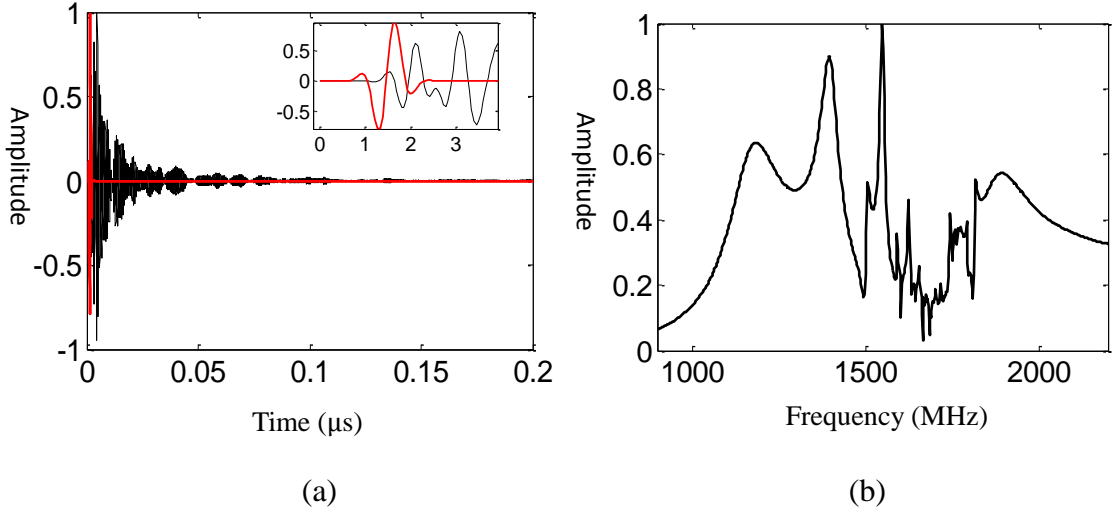

(a)                                                              (b)

FIG. 4 (a) A small dipole source of 2.8ns (red line), which is placed at z=-0.1$\lambda_0$ away from the bottom of the resonant aperture ($\lambda_0 = 188mm$). The response at $\boldsymbol{r}_d = (0, 94mm, 62mm)$ is recorded and shown by the black line. The inset is the zoomed part. (b) The normalized amplitude of the frequency-dependent response at $\boldsymbol{r}_d$. This set of figures shows many abrupt changes within a very small frequency separation, implying that the response is highly sensitivity to frequency.

Equation (4) shows that $|\frac{d}{d\omega}\mathbf{E}(\boldsymbol{r}_d;\boldsymbol{k}_{in})|$ exhibits peaks as a function of frequency $\omega$ for a given $k_\rho > k_0$ subject to $R_{op}^{TM}e^{ik_{pz}d} = \pm 1$, which implies that the frequency-dependent response exhibits a large amount of resonant peaks within a certain frequency range when a broadband dipole source is used. We perform another set of numerical simulations using the commercial software of CST Microwave Studio 2012. We excite the resonant aperture with a small *x*-polarized dipole source located 18 nm from the bottom of resonant aperture, where this source emits a Gaussian pulse with 2.5ns duration (see the red line in Fig. 4(a)), and both its frequency-dependent and time-dependent responses are acquired at $\boldsymbol{r}_d$ (see black lines in 4(a) and 4(b), respectively). This set of figures demonstrates that a broadband pulse will show rich peaks in the frequency domain after experiencing the resonant aperture; accordingly, it will be fully expanded in the time domain with a factor of more than 100. Consequently, from the respect of the information capacity, the degree of freedom of the measurements will be considerably driven up with a factor of 100 and beyond.

From the viewpoint of signal sampling, the high sensitivity to frequency implies that the sampling space on frequencies should be sufficiently small, and thus is related to the increase of the degree of freedom (DoF) of measurements in the far-field region. In this way, the information capacity of the imaging system is increased, which accounts for the enhanced imaging resolution. We also emphasize that the near-field coupling between the imaged objects and resonant aperture encodes subwavelength details of objects into the temporal measurements in the far-field region. Hence, in order to obtain the subwavelength imaging from far-field measurements by a single broadband radar, we conclude that three basic requirements should be satisfied: the strong spatial-temporal resonant aperture antenna, near-field coupling, and temporal (or broadband) illumination.

Based on above principles, we propose the single-radar far-field imaging beyond the diffraction limit, which is verified by full-wave numerical simulations. Here, the simulation parameters are set as: $\mathbf{r}_d = (5\lambda_0, 5\lambda_0, 10\lambda_0)$, and the operational wavelength varies from 150 mm to 200 mm with a step of 0.002 mm. In addition, the object under consideration is placed at $0.05\lambda_0$ from the bottom of the resonant aperture. The probed object consists of a Chinese character "北", with a refraction index of 2.7, as illustrated in Fig.5(a). The distance between the centers of the two neighbored objects is set to be 12 mm. For this setup, we voluntarily opt for a low-refractive-index-contrast, which is typical for soft-matter objects. In addition, the simulation data input of full-vectorial electric field to the reconstruction procedure is generated by using CST Microwave Studio 2012.

Basically, the frequency-dependent measurement data $E(\omega)$ acquired at $r_d$, is related to the probed object characterized by its contrast $O(r)$ by $E(\omega) = \int_D \boldsymbol{G}(\boldsymbol{r}_d, \boldsymbol{r}'; \omega) E_{\text{in}}(\boldsymbol{r}'; \omega) O(\boldsymbol{r}') d\boldsymbol{r}'$, where $E_{\text{in}}(\boldsymbol{r}'; \omega)$ is the incidence field for the presence of the resonant aperture, and the integral is performed over the region of interest. Herein, $\boldsymbol{G}(\boldsymbol{r}_d, \boldsymbol{r}'; \omega)$ denotes the Green's function relating a point source at $\boldsymbol{r}'$ and its response at $\boldsymbol{r}_d$. Note that both $\boldsymbol{G}(\boldsymbol{r}_d, \boldsymbol{r}'; \omega)$ and $E_{\text{in}}(\boldsymbol{r}'; \omega)$ are known to be a *prior*. Regarding the inverse procedure, the object is retrieved from the frequency-dependent measurements $E(\omega)$ by optimizing the following cost function, i.e.,

$$\hat{O}(\mathbf{r}) = \text{argmin}_{O(\mathbf{r}')} \left[ \int \left| E(\omega) - \int_D \mathbf{G}(\mathbf{r}_d, \mathbf{r}'; \omega) E_{\text{in}}(\mathbf{r}'; \omega) O(\mathbf{r}') d\mathbf{r}' \right|^2 d\omega + \gamma \int_D |O(\mathbf{r}')| d\mathbf{r}' \right] \quad (5)$$

In Eq. (5), the second term involved in the optimized object function stands for the penalty term to stabilize the optimize procedure, and $\gamma$ is a balance factor. The reconstruction of solving Eq. (5) is achieved by using the iteratively reweighed approach [18]. The reconstructed results are shown in Fig. 5(b), where the additive Gaussian noise with different noise level 20dB, 30dB, and 40dB SNR have been added to the simulated data. To investigate the importance of the resonant aperture for the far-field imaging beyond the diffraction limit, we perform simulations with parameters similar to above but without the resonant aperture and the associated results are shown in Fig. 5(c). To underline the importance of broadband illumination, we conduct the reconstruction as illustrated in Fig. 5(d), where $50 \times 50$ receivers are uniformly distributed over a square of $10\lambda_0 \times 10\lambda_0$ at $z=10\lambda_0$. From Figs.5, we clearly see that the necessity of the three basic requirements for the subwavelength imaging, the resolution of about $\lambda_0/20$ from far-field measurements. Despite the simplicity of the imaged objects, results prove that the subwavelength information of an object, which is registered in the far-field region by the spatio-temporal resonant lens, can be restored by processing the temporal data acquired by using single radar. We remark that our approach is at the frontier of what is achievable with current experimental techniques.

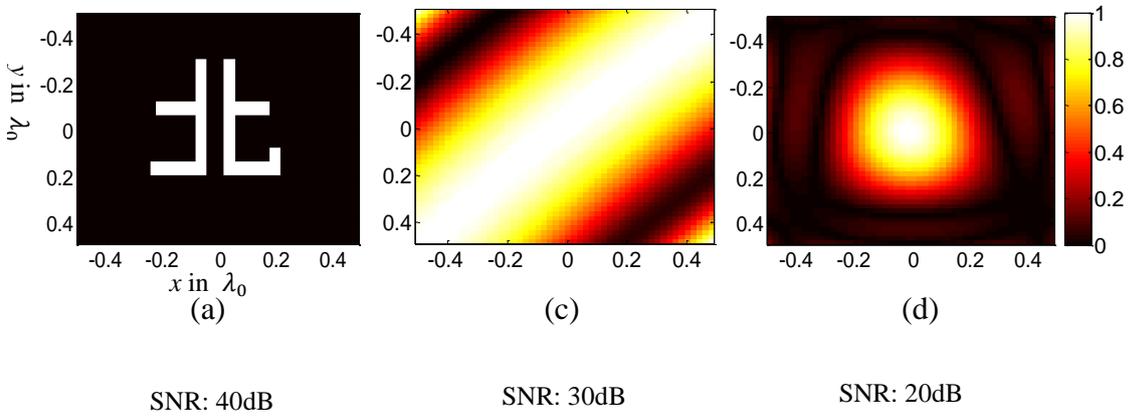

(a) SNR: 40dB     (c) SNR: 30dB     (d) SNR: 20dB

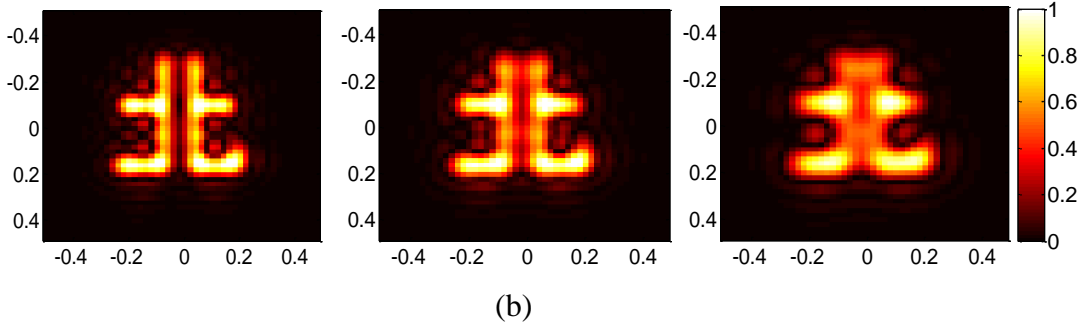

(b)

FIG. 5 Ground truth and reconstructed results by different methods. (a)The ground truth consisting of a Chinese character "北". (b)Reconstruction results for different noise level of 40dB, 30dB, and 20dB using a single radar imaging system as sketched in Fig. 1(a). In the simulation, the master antenna is located at $\mathbf{r}_d = (5\lambda_0, 5\lambda_0, 10\lambda_0)$, the wavelengths used for broadband illumination varies from 150mm to 300 mm with a step of 0.002 mm, and the probed objects is located at $0.05\lambda_0$ at the bottom of the resonant aperture antenna. (c)Reconstruction results without the resonant aperture antenna, where other simulation parameters are the same as those used in Fig. 5(b) but SNR being 30dB. (d)Reconstruction result without the resonant aperture antenna, where 50× 50 receivers are uniformly distributed over a square of $10\lambda_0 \times 10\lambda_0$ at z=$10\lambda_0$, and SNR is set to be 30dB.

**Conclusion.** In summary, we develop a concept of single-radar system for super-resolution imaging, which benefits from the use of spatial-temporal aperture and sparse reconstruction. We have shown that for a given finite operational bandwidth, the resonant-aperture antenna can increase the information capacity of measurements efficiently in far fields due to strong Fabry-Perot resonance, leading to the far-field imaging beyond the diffraction limit. We have theoretically demonstrated that three basic requirements should be satisfied to obtain the subwavelength imaging from far-field measurements: the spatial-temporal resonant aperture antenna, near-field coupling, and broadband illumination. A set of proof-of-concept investigations has been performed to verify the proposed theory. It is expected that such a single-radar concept can find real applications for subwavelength imaging by using more specialized spatial-temporal aperture and efficient reconstruction solver. In addition, the proposed single-radar imaging system amounts to encode the spatial details of the probed targets into temporal domain, which mathematically leads to a measurement matrix well matched to CS, and serves as an easy-implementation apparatus for compressive measurements [19, 20].

$T^{TM}e^{ik_{iz}d}e^{i\mathbf{k}_{inc,\perp}\cdot\mathbf{r}_\perp}$ for $\mathbf{r}_\perp \in \Omega$, otherwise being zero. Of course, more accurate analysis can be made by via a full-wave solver of the coupled-mode method, for instance,

Appendix A:

Assuming that the resonant aperture is located at *z*=0, and its normal direction is parallel to the *z*-axis, as sketched in Fig. 1. In light of the Huygens's theorem [7], the diffracted field arising from the electrical field at the resonant aperture, $\mathbf{E}_t(\boldsymbol{\rho}', 0)$, is express as

$$\mathbf{E}(\boldsymbol{\rho}, z) = 2\nabla \times \int [\hat{\mathbf{z}} \times \mathbf{E}_t(\boldsymbol{\rho}', 0)] G(\boldsymbol{\rho}-\boldsymbol{\rho}', z) d\boldsymbol{\rho}' \qquad (S1)$$

Herein, $\boldsymbol{\rho} = [x\ y]^T$, $d\boldsymbol{\rho}' = dx'dy'$, $G(\boldsymbol{\rho}-\boldsymbol{\rho}',z)$ is the three dimensional scalar Green's function in free space. Note that above integral is performed over the radiation aperture $\Omega$, which implies that $\mathbf{E}_t(\boldsymbol{\rho}',0) = 0$ for $\boldsymbol{\rho}' \notin \Omega$. Introducing the far-field approximation of $G(\boldsymbol{\rho}-\boldsymbol{\rho}',z) \approx \frac{e^{ik_0 r_d}}{4\pi r_d} e^{-i\boldsymbol{k}_r \cdot \boldsymbol{\rho}'}$ in Eq.(S1), we arrive immediately at

$$\mathbf{E}(\boldsymbol{\rho},z) = 2\nabla \times \frac{e^{ik_0 r_d}}{4\pi r_d} \int [\hat{\mathbf{z}} \times \mathbf{E}_t(\boldsymbol{\rho}',0)] e^{-i\boldsymbol{k}_{r,\rho} \cdot \boldsymbol{\rho}'} d\boldsymbol{\rho}' \tag{S2}$$

with $\boldsymbol{k}_r = k_0(\sin\theta_r\cos\varphi_r, \sin\theta_r\sin\varphi_r, \cos\varphi_r)$, $\theta_r$ and $\varphi_r$ are the polar angle and azimuth angle, respectively. Using the specific formula for $\mathbf{E}_t(\boldsymbol{\rho}',0)$ as discussion below, and taking $\hat{\mathbf{z}} \times \mathbf{E}_t(\boldsymbol{\rho}',0) = \boldsymbol{C}T^{TE,TM}e^{-i\boldsymbol{k}_{inc,\perp}\cdot\boldsymbol{\rho}'}$ into account, we can simplify Eq.(S2) into

$$\mathbf{E}(\boldsymbol{\rho},z) = 2\nabla \times \left(\frac{e^{ik_0 r_d}}{4\pi r_d} \boldsymbol{C}T^{TE,TM} \int e^{i(\boldsymbol{k}_{inc,\rho} - \boldsymbol{k}_{r,\rho})\cdot\boldsymbol{\rho}'} d\boldsymbol{\rho}'\right)$$

$$= 2L_x L_y T^{TE,TM} \operatorname{sinc}\left(\frac{L_x(k_{r,x}-k_{in,x})}{2\pi}\right) \operatorname{sinc}\left(\frac{L_y(k_{r,y}-k_{in,y})}{2\pi}\right) \nabla \times \left(\boldsymbol{C}\frac{e^{ik_0 r_d}}{4\pi r_d}\right) \tag{S3}$$

where $\boldsymbol{C}$ is a constant vector, $T^{TE,TM}$ are dependent on the transmission coefficients associated with specific TE or TM illumination polarizations.

For this TM illumination, the incident plane wave is

$$\boldsymbol{E}_{inc} = (\hat{\boldsymbol{x}}\cos\theta_i\cos\varphi_i + \hat{\boldsymbol{y}}\cos\theta_i\sin\varphi_i - \hat{\boldsymbol{z}}\sin\theta_i)e^{ik_{inc,z}z}e^{i\boldsymbol{k}_{inc,\rho}\cdot\boldsymbol{\rho}} \tag{S4}$$

Then, the transversal field at the output of the resonant aperture is

$$\boldsymbol{E}_t = \frac{k_{inc,z}}{k_{in,\rho}^2}\sin\theta_i T^{TM} e^{ik_{in,z}z}\boldsymbol{k}_{in,\rho}e^{i\boldsymbol{k}_{in,\rho}\cdot\boldsymbol{\rho}}, \quad \boldsymbol{\rho}\in\Omega \tag{S5}$$

$$= 0, \quad \boldsymbol{\rho}\notin\Omega$$

where

$$T^{TM} = C_p e^{ik_{pz}d} + D_p e^{-ik_{pz}d}, \tag{S6}$$

$$D_p = -\frac{R_{op}^{TM}T_{op}^{TM}e^{i2k_{pz}d}}{1-(R_{op}^{TM})^2 e^{i2k_{pz}d}} \tag{S7}$$

$$C_p = \frac{T_{op}^{TM}}{1-(R_{op}^{TM})^2 e^{i2k_{pz}d}} \tag{S8}$$

$$R_{op}^{TM} = \frac{k_{in,z}\varepsilon_r - k_{pz}}{k_{in,z}\varepsilon_r + k_{pz}} \tag{S9}$$

$$T_{op}^{TM} = \frac{2k_{in,z}\varepsilon_r}{k_{in,z}\varepsilon_r + k_{pz}} \tag{S10}$$

For completeness, we also give corresponding results for the case of TE illumination. For this case, the incident plane wave is

$$\boldsymbol{E}_{inc} = (-\hat{\boldsymbol{x}}\sin\varphi_i + \hat{\boldsymbol{y}}\cos\varphi_i)e^{ik_{in,z}z}e^{i\boldsymbol{k}_{in,\rho}\cdot\boldsymbol{r}_\perp} \tag{S11}$$

Then, the transmission electrical field is

$$\mathbf{E}_t(\boldsymbol{\rho}, 0) = \frac{\omega\mu_0}{k_{in,\rho}^2}\frac{1}{\eta_0}E_{hi}\sin\theta_i T^{TE}e^{-jk_{iz}z}\boldsymbol{k}_{in,\rho}\times\hat{\boldsymbol{z}}e^{-j\boldsymbol{k}_{in,\rho}\cdot\boldsymbol{\rho}}, \quad \boldsymbol{\rho}\in\Omega \tag{S12}$$

$$= 0, \quad \boldsymbol{\rho}\notin\Omega$$

where

$$T^{TE} = A_p e^{ik_{pz}d} + B_p e^{-ik_{pz}d} \tag{S13}$$

$$A_p = \frac{T_{op}^{TE}}{1-(R_{op}^{TE})^2 e^{i2k_{pz}d}} \tag{S14}$$

$$B_p = -\frac{T_{op}^{TE}R_{op}^{TE}e^{i2k_{pz}d}}{1-(R_{op}^{TE})^2 e^{i2k_{pz}d}} \tag{S15}$$

$$R_{op}^{TE} = \frac{k_{in,z}-k_{pz}}{k_{in,z}+k_{pz}} \tag{S16}$$

$$T_{op}^{TE} = \frac{2k_{in,z}}{k_{in,z}+k_{pz}} \tag{S17}$$